# 2D mapping of radiation dose and clonogenic survival for accurate assessment of *in vitro* X-ray GRID irradiation effects

Delmon Arous[1,2], Jacob Larsen Lie[1], Bjørg Vårli Håland[1], Magnus Børsting[1], Nina Frederike Jeppesen Edin[1] and Eirik Malinen[1,2]

[1] Department of Physics, University of Oslo, PO Box 1048 Blindern, N-0316, Oslo, Norway
[2] Department of Medical Physics, The Norwegian Radium Hospital, Oslo University Hospital, PO Box 4953 Nydalen, N-0424 Oslo, Norway

E-mail: delmon.arous@fys.uio.no

## Abstract

Spatially fractionated radiation therapy (SFRT or GRID) is an approach to deliver high local radiation doses in an 'on-off' pattern. To better appraise the radiobiological effects from GRID, a framework to link local radiation dose to clonogenic survival needs to be developed. A549 (lung) cancer cells were irradiated in T25 cm$^2$ flasks using 220 kV X-rays with an open field or through a tungsten GRID collimator with periodical 5 mm openings and 10 mm blockings. Delivered nominal doses were 2, 5, and 10 Gy. A novel approach for image segmentation was used to locate the centroid of surviving colonies in scanned images of the cell flasks. Gafchromic$^{TM}$ film dosimetry (GFD) and FLUKA Monte Carlo (MC) simulations were employed to map the dose distribution in the flasks at each surviving colony centroid. Fitting the linear-quadratic (LQ) function to clonogenic survival data for open field irradiation, the expected survival level at a given dose level was calculated. The expected survival level was then mapped together with the observed levels in the GRID-irradiated flasks. GFD and FLUKA MC gave similar dose distributions, with a mean peak-to-valley dose ratio of about 5. LQ-parameters for open field irradiation gave $\alpha = 0.16 \pm 0.04$ Gy$^{-1}$ and $\beta = 0.001 \pm 0.004$ Gy$^{-2}$. Using the image segmentation method, the mean absolute percentage deviation between observed and predicted survival in the (peak; valley) dose regions was (8; 10) %, (4; 41) %, and (3; 138) % for 2, 5 and 10 Gy, respectively. In conclusion, a framework for mapping of surviving colonies following GRID irradiation together with predicted survival levels from homogeneous irradiation was presented. For the given cell line, our findings indicate that GRID irradiation, especially at high peak doses, causes reduced survival compared to an open field configuration.

## 1. Introduction

In conventional radiation therapy (CRT), the treatment aim is to ensure that the whole tumor receives a homogeneous radiation dose. Spatially fractionated radiation therapy (SFRT) was originally introduced in 1909 by Köhler, where irradiation was performed through a perforated screen to deliver high radiation doses and, at the same time, to reduce extensive toxicity of skin and subcutaneous tissue (Laissue *et al.*, 2011). In SFRT, an 'on-off' field pattern is formed using an array of narrow parallel beamlets that are periodically spaced by a fixed gap, laterally producing regions of low (valley) and high (peak) dose. Therefore, SFRT delivers a heterogenous dose distribution to the tumor, in contrast to CRT employing a homogeneous dose. SFRT has so far mainly been employed with a palliative treatment intent, and for this dose peaks higher than 10 Gy per fraction are commonly delivered. Although SFRT delivery may be achieved using fixed collimators of dense metals or metal alloys such as Cerrobend, other solutions can also be utilized such as multileaf collimators (MLCs) (Ha *et al.*, 2006).

GRID is a two-dimensional (2D) treatment technique of SFRT, in which the beamlet size is typically in the range of 7.5-15 mm. The dose is normally delivered as a single radiation fraction (10-20 Gy peak dose) in a systematic pattern. GRID has been used both as monotherapy and in combination with external CRT and chemotherapy to enhance the local control of large and bulky tumors (Mohiuddin *et al.*, 1999; Huhn *et al.*, 2006; Peñagarícano *et al.*, 2010). Because of the small and discrete areas of irradiated healthy tissue by this method, higher doses may be tolerated than by homogeneous irradiation. Published clinical evidence from GRID therapy studies showed beneficial palliative outcomes with reduced toxicity (Mohiuddin *et al.*, 1990; Mohiuddin *et al.*, 1996; Neuner *et al.*, 2012).

The primary aim of this current investigation was to develop appropriate tools that allows to map both the dose deposition and the clonogenic survival following GRID irradiation *in vitro*. It is necessary to establish such a methodology to accurately appraise the radiobiological impact and to understand the subsequent biological responses. Our proposed methodology consists principally of two parts: 1) film dosimetry to map the resulting dose distribution in the cell flasks and 2) mapping of surviving colonies. The dosimetry was done by Gafchromic™ films together with Monte Carlo (MC) simulations. Importantly, dosimetry and colony scoring were done in the same spatial frame of reference. Furthermore, the linear-quadratic (LQ) cell survival model was fitted to clonogenic survival following homogenous irradiation. Applying the model with the GRID dose distribution will give the expected survival pattern if the cellular response was identical to that for homogenous irradiation. This enabled a comparison of the observed clonogenic survival following GRID irradiation to the survival predicted from open field irradiation. The developed framework is expected to facilitate high-quality studies of GRID effects and contribute to better insight into the differential mechanisms following homogenous and heterogeneous dose deliveries.

## 2. Methods

### 2.1 Cell culture and clonogenic survival assays

Human non-small cell lung cancer (NSCLC) A549 cells (Giard *et al.*, 1973) were grown in a BioWhittaker 1:1 mix of Dulbecco's Modified Eagle Medium (DMEM) and F12 with 15 mM Hepes and L-Glutamine, supplemented with 10% of fetal bovine serum (FBS) (Biochrom), 1% penicillin/streptomycin (Lonza) and 200 units per liter insulin (Gibco). The cells were maintained in a fully humidified incubator providing 5% $CO_2$ at 37°C. This cell line was chosen because of its ability to form well-defined colonies.

Before irradiation, the cells were seeded into T25 $cm^2$ flasks (Nunclon, Denmark) in 5 mL culture medium at density of 30 000 cells per flask to ensure appropriate colony scoring over a wide range of survival levels. To exclude artifacts from biological processes affected by cell density, the number of cells seeded was the same for all experiments. This cell density was chosen as high enough to be able to measure the effects of dose gradients in the penumbra regions of the grid, but low enough for detection of individual colonies. The plating efficiency (PE) observed for the A549 cell line was 7-12% in our experiments.

All flasks were incubated overnight for 24 h prior to irradiation, allowing cells to adhere. After irradiation, the cell flasks were incubated for about 6 days before the colonies were fixated in 96% technical ethanol (Antibac, Norway) and stained with Methylene Blue (Sigma, USA).

The cell flasks were then scanned in order to quantify the clonogenic survival using an in-house machine learning colony segmentation method (see Section 2.4). Clonogenic survival was chosen as the endpoint to assess the effects of radiation, where survival was defined as the ability of a single cell to grow into a colony composed of more than 50 cells (Franken *et al.*, 2006).

### 2.2 Gafchromic™ film dosimetry (GFD)

GFD was used to measure the 2D dose distribution in the T25 cell culture flasks. The dosimetry experiment was conducted separately from the irradiation of the A549 cells. Gafchromic™ EBT3 films (lot No. 02122001) were handled and processed according to the recommended protocols specified for radiochromic film dosimetry in the report of AAPM Task Group 235 (Niroomand-Rad *et al.*, 2020).

Prior to irradiation, the EBT3 films were cut in dimensions to fit inside T25 flasks (6.5×4.3 $cm^2$), where the upper right corner was cut to mark the film orientation. Films were irradiated inside the flasks with open and GRID fields (see below). Using the same batch of EBT3 films, irradiation of dose calibration films was carried out prior to following film dose measurements and was conducted with delivered doses of $D = 0, 0.1, 0.2, 0.5, 1, 2, 5$ and $10$ Gy. In order to reduce the statistical error, each calibration point consisted of eight irradiated film pieces (Devic *et al.*, 2006).

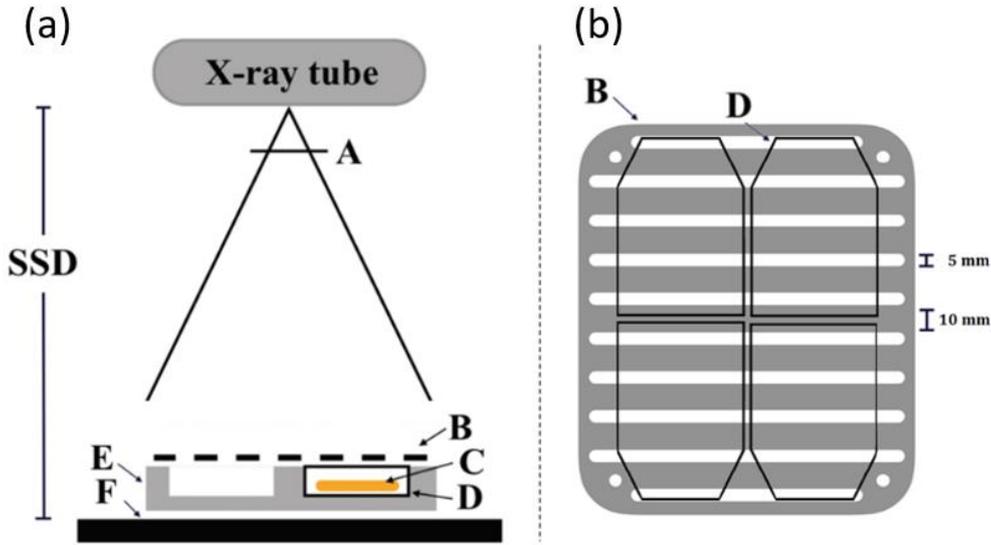

**Figure 1**. (*a*) Experimental setup for GRID irradiation: X-ray filtration (**A**), tungsten grid collimator (**B**), Gafchromic™ EBT3 film/A549 cell suspension (**C**), T25 cm2 cell culture flask (**D**), PMMA cell flask holder (**E**), Perspex table (**F**). (*b*) Beam's eye view of the grid collimator with underneath placement of T25 flasks outlined.

An Epson Perfection V850 Pro flatbed scanner (Epson, Japan) and its associated software EPSON Scan v3.9.3.3 were used to read all films 48 h after irradiation. To minimize film response uncertainty and variation due to film orientation dependence, nonuniformity and lack of reproducibility in the scanner response, all EBT3 films were scanned in the same orientation, positioned at a reproducible central location of the scan surface that was considered uniform. Each film was scanned four consecutive times. The digitized images were acquired in transmission and RGB-positive mode with 16-bit depth per color channel and 300 dots per inch (dpi) spatial resolution, and without applying any image adjustments or color corrections. The images were saved in .tiff format.

Further image processing and analysis of the digitized images were done with MATLAB R2020b (MathWorks, Natick, MA, USA). The images were processed in three colors channels (red, green and blue) and as grayscale image, where a predefined region of interest (ROI) of $4\times4$ mm$^2$ size (Gholizadeh Sendani *et al.*, 2018) was selected from the center of each calibration film to obtain the mean and standard deviation of the transmitted light intensity pixel value (PV). Utilizing an in-house made MATLAB script, the sampled mean PV was used to characterize the film response by the net optical density (netOD) with background correction (Devic *et al.*, 2016):

$$netOD = \log_{10}\left(\frac{PV_{ctrl} - PV_{bckg}}{PV_{rad} - PV_{bckg}}\right), \qquad (1)$$

where $PV_{ctrl}$, $PV_{rad}$ and $PV_{bckg}$ are the averaged PV from images of control (unirradiated) EBT3 films, irradiated EBT3 films and "absolutely" opaque sheets, respectively.

Furthermore, an appropriate analytical function was chosen to establish the calibration relationship between the netOD values and the known absorbed doses (Devic *et al.*, 2004):

$$D = a \cdot netOD + b \cdot netOD^n,$$

where $n$ initially was iteratively varied from 0.5 to 5.0 with a step of 0.5, because treating $n$ as a fitting parameter introduces a higher overall fit uncertainty while only negligibly improving of the sum of residuals (Devic *et al.*, 2004). The goodness of the least-squares curve fitting was measured by R-squared ($R^2 = 0.999$ for red color channel selection).

### 2.3 GRID and open irradiation

Either GRID (heterogenous) or open (homogeneous) field irradiation of A549 lung cancer cells and radiochromic EBT3 films was performed using a Pantak PMC 1000 X-ray unit (Pantak, USA). The tube was operated at 220 kV and 10 mA with a 0.70 mm Cu and a 1.52 mm Al filter to yield a dose rate of 0.59 Gy/min. The cell clonogenicity and dosimetry experiments were conducted at the same lateral position and source-surface distance (SSD) of 60 cm (see Figure 1). GRID irradiation was done using a custom-made tungsten collimator with periodical 5 mm openings and 10 mm blockings creating a striped irradiation field. It was formed to fit and rest on a polymethyl methacrylate (PMMA) cell flask holder, which was in-house designed and built with four cavities to contain the T25 cell culture flasks. Both the cells and radiochromic films were irradiated in T25 flasks, which were placed in the PMMA cell flask holder. For cell irradiation, the PMMA holder was placed on a preheated PMMA plate maintaining 37°C in the medium of the flasks by circulating air.

For the A549 cell colonies, the delivered single doses were 0, 2, 5, and 10 Gy with 4 replicates per dose per field pattern in each of three experiments. The reported doses are actual doses for the open field configuration, while they serve as nominal doses for the GRID dose distributions. For the EBT3 measurement films, single dose of 5 Gy was delivered to 16 film pieces per field pattern.

*2.4 Colony image segmentation*

A novel machine learning approach using principal component-based watershed method for image segmentation was developed to identify individual colonies in a cell flask (Arous *et al.*, 2022). This automated algorithm was applied to scanned images of the cell culture flasks containing fixed and stained A549 cell colonies to locate the centroid of viable colonies. The images were processed with predefined parameters for the current cell line, where conglomerations composed of more than 50 cells were counted as a colony. Thereby, spatial 2D cell colony survival distributions were produced.

The image data were obtained from the flatbed scanner specified above for radiochromic dosimetry. The scanner provided RGB- images in transmission mode with 16-bit depth per color channel and 1200 dpi spatial resolution. For consistency, the cell flasks were positioned at the same location of the scan surface as the scanned EBT3 films. Moreover, no prior filtering nor adjustments were performed on the captured images in conjunction with the scanning.

*2.5 FLUKA Monte Carlo (MC)*

MC simulations were performed using FLUKA code version 4-2.1 (Böhlen *et al.*, 2014; Battistoni *et al.*, 2015) and its graphical user interface Flair version 3.1-15 (Vlachoudis, 2009). The simulations were used to model the experimental irradiation setup and to calculate the theoretical dose distribution in the region of the T25 flasks containing the EBT3 films (or analogously, the cell colonies). This served as a means to verify the GFD following open and GRID irradiation.

In the code, the focal spot of the X-ray unit along with the propagation of the primary X-ray photons to the EBT3 films contained in the T25 flasks was simulated, where the material geometry of the entire irradiation setup was modelled as described in Section 2.3. The layer structure and corresponding elemental composition of the EBT3 films encased inside the flasks was also included in the modelling (Devic *et al.*, 2016). A 220 kV X-ray spectrum sequentially attenuated through a 0.70 mm Cu and a 1.52 mm Al filter was simulated. The spatial shape of the beam imitated the focal spot with a Gaussian lateral spread of 0.2 mm FWHM. Furthermore, the MC acquisition was performed by adopting default FLUKA physics settings (cf. PRECISIOn defaults in the FLUKA manual (Ferrari *et al.*, 2005)), where transport and production cut-off of photons and secondary electrons was set to 1 keV. In total, $2.0\times10^7$ primary X-ray photons were simulated.

Moreover, four scoring regions were defined for each EBT3 film piece within the flasks. Both dose deposition and fluence were obtained in the scoring regions by using the USRBIN option in FLUKA. Here, the quantities were scored in a spatial regular mesh of $732\times507\times1$ uniformly sized bins enveloping the dimensions of the modelled EBT3 films of $6.5\times4.3\times0.0278$ cm$^3$. The mesh resolution of the lateral scoring area was chosen to match the spatial resolution of the digitized EBT3 images for subsequent comparison.

*2.6 Observation and prediction of local clonogenic survival*

A direct spatial comparison of local colony survival distribution after GRID irradiation (observed survival) and the calculated survival for the measured GRID doses using parameters from a fit by the generalized LQ model (Fowler, 1989) to data from open field irradiation (predicted survival) was performed (see Figure 2).

From the cell colony segmentation algorithm, the centroid of each detected viable colony defined the experimental colony survival maps in the cell culture flasks. In all cell colony irradiations, the surviving fraction (SF) was normalized to the PE of control flasks (i.e., sham treated and time-matched incubation flasks):

$$SF|_{\Delta x} = \frac{N_{count}}{N_{seed} \cdot PE}|_{\Delta x}, \qquad (2)$$

where $\Delta x = 1\ mm$ is the band thickness longitudinally across the cell flask, $N_{count}|_{\Delta x}$ is the number of identified colony centroids within $\Delta x$ made by the segmentation algorithm, $N_{seed}$ is the number of plated cells in all experiments and $PE$ is the plating efficiency for the unirradiated A549 cells. The observed survival map was obtained by calculating the number of viable colony centroids within each $\Delta x$ for each cell flask and then average for each $\Delta x$ over all cell flasks. From this, 95% confidence intervals (CIs) in experimental cell colony survival were obtained.

The dose map obtained from GFD was then used to estimate the dose at each surviving colony centroid. The predicted local SF was calculated from the LQ model fitted to data from open field irradiation:

$$SF(D(r)) = e^{-(\alpha D(r) + \beta D^2(r))}, \qquad (3)$$

where $r$ is the lateral position with coordinate $(x, y)$, $D(r)$ is the local dose described by the 2D EBT3 dose distribution, $\alpha$ and $\beta$ are estimated by linear regression for A549 SF following open field irradiation with single nominal doses of $D = 0, 2, 5$ and $10$ Gy. Similarly as for GRID-irradiated colonies, the predicted survival level in each pixel was mapped together with observed dose levels for single nominal doses of $D = 0, 2, 5$ and $10$ Gy. Corresponding 95% CIs were also estimated for open and GRID field pattern.

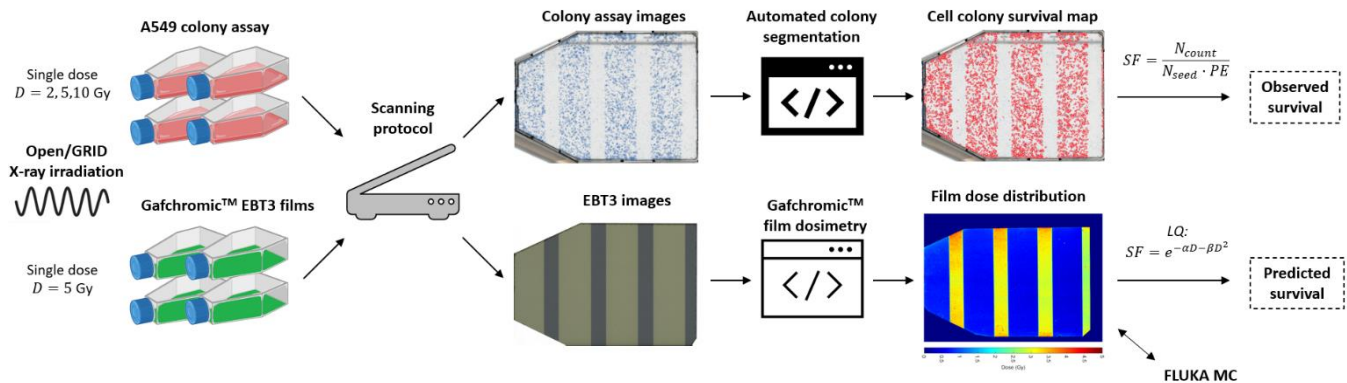

**Figure 2**. Overview of the data acquisition and assessment processing pipeline showing the main steps. Open or GRID irradiation with 220 kV X-ray beam was performed with a single dose $D$ of A549 colonies and radiochromic EBT3 films in T25 cm$^2$ cell culture flasks. Subsequently, colony formation assays and radiochromic EBT3 films were scanned, where an in-house principal component-based watershed method (Arous *et al.*, 2022) for image segmentation was used to locate the centroid of each surviving colony. Gafchromic$^{TM}$ film dosimetry (GFD) was performed to measure the two-dimensional (2D) dose distribution in the cell flasks. FLUKA Monte Carlo simulation imitating the GRID irradiation experiments was done to verify the 2D GFD. From surviving fraction (SF) following open field irradiation, α and β in the linear-quadratic (LQ) survival model was estimated by linear regression. Then, the average predicted survival levels in each pixel were then mapped together with the average observed levels in the GRID-irradiated flasks.

## 3. Results

### 3.1 Dosimetry validation

FLUKA MC simulation imitating the open and striped GRID irradiation experiments was performed to verify the GFD. Utilizing the FLUKA code, the experimental setup was modelled and the dose distributions delivered to the EBT3 films contained in the cell culture flasks were calculated.

Having performed the GFD and scored the dose distributions with MC, central profiles were obtained across these 2D dose maps for validation. To scale these maps for comparison, they were normalized with respect to respective averaged dose profiles for open irradiation (see Figure 3). When exposing the films to a nominal 5 Gy dose, the films exposed to an open field are exhibiting a mean 5 Gy response, whereas using the striped GRID, the average valley dose was found to be 0.9 Gy and the average peak dose to be 4.1 Gy (see Table 1). Hence with GRID irradiation, about 18% of the dose deposition is lost in the peak regions when compared to open field irradiation. The GFD and MC simulations gave highly similar dose distributions, with a mean peak-to-valley dose ratio (PVDR) of 4.7 (4.5, 4.9) and 5.2 estimated from the EBT3 dosimetry and FLUKA MC simulations, respectively.

**Table 1**. Dose estimates from EBT3 films following open and GRID irradiation are presented together with 95% CIs estimated from the Gafchromic$^{TM}$ film dosimetry (GFD).

| Open | GRID | |
|---|---|---|
| Dose (Gy) (95% CI) | Valley dose (Gy) (95% CI) | Peak dose (Gy) (95% CI) |
| 5.0 (4.8, 5.1) | 0.9 (0.8, 0.9) | 4.1 (3.9, 4.2) |

### 3.2 Local and average survival comparison after GRID irradiation

Fitting the LQ model to clonogenic dose response measurements of A549 cells for open field irradiation by linear regression (see Figure 4) gave parameter values of $\alpha = 0.16 \pm 0.04$ Gy$^{-1}$ ($p$-value $< 0.05$) and $\beta = 0.001 \pm 0.004$ Gy$^{-2}$ ($p$-value $> 0.05$). The goodness-of-fit measure was $R^2 = 0.673$.

Having established the values of α and β, the LQ model from equation (2) was deployed on the measured EBT3 dose distributions to produce 2D maps of the predicted SF (see Figure 5, middle panel). The experimental SF was found using the image segmentation method, in which the centroid of each surviving cell colony was located (see Figure 5, left panel). Equation (3) was then employed to map the average observed SF profiles in the cell flask to compare with the predicted profiles as shown in Figure 5 (right panel) for three nominal doses (2, 5 and 10 Gy). Qualitatively for the A549 cell line, mapping the SF longitudinally along the cell flask gave a pattern resembling the GRID collimator outline, where the average observed and predicted survival profiles also gave similar patterns. However, significantly lower observed survival was found in the peak regions compared to predicted survival. A similar, but not significant, tendency was seen in the valley regions for 2 Gy nominal dose, while a proliferative tendency was observed for 10 Gy nominal dose.

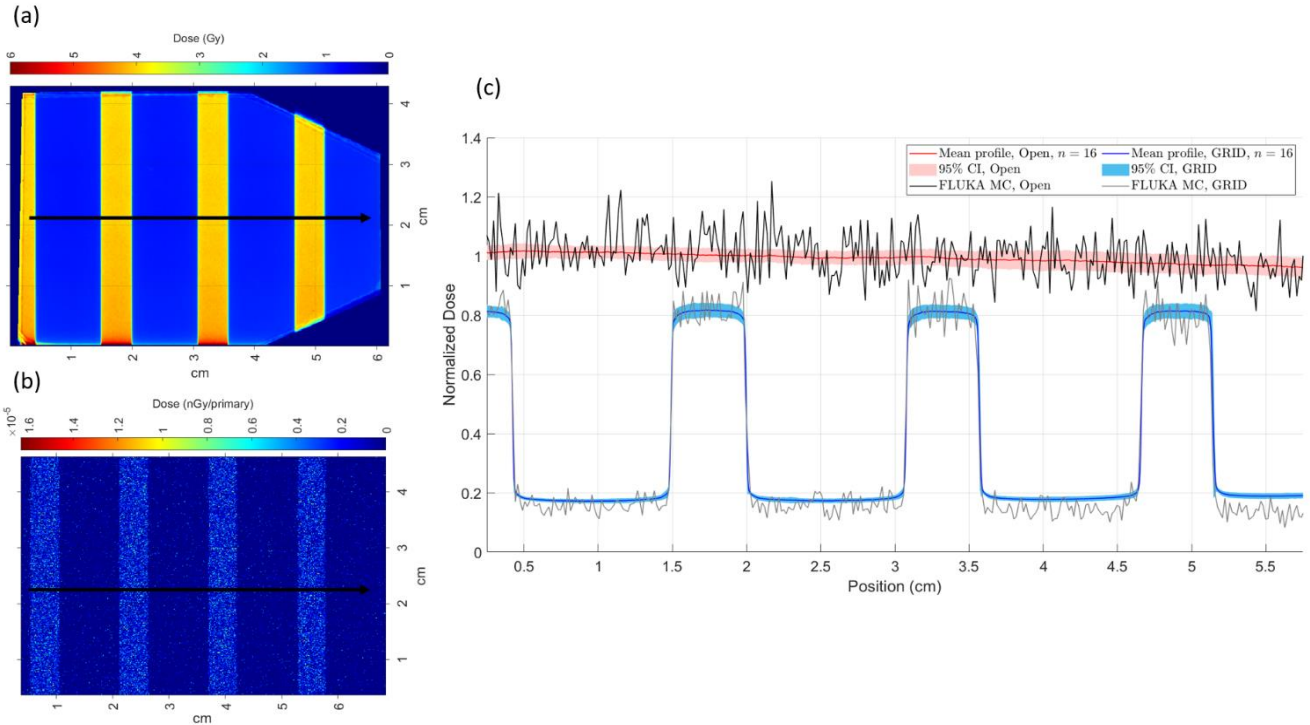

**Figure 3**. (*a*) Two-dimensional (2D) dose distribution for GRID irradiation measured by EBT3 Gafchromic™ film dosimetry (GFD). (*b*) 2D dose distribution scored with FLUKA Monte Carlo (MC) simulations. (*c*) Normalized dose profiles from EBT3 GFD of open and GRID irradiation together with normalized dose profiles from FLUKA MC simulations. Dose profiles from EBT3 film dosimetry for open and GRID irradiation are presented with red and blue lines, respectively. Correspondingly coloured 95% CIs for $n = 16$ radiochromic EBT3 films per field configuration are presented with bands. Dose profiles from MC simulations for open and GRID irradiation are presented with black and grey lines, respectively.

Quantitatively, the absolute mean percentage deviations between the observed and predicted survival in valley and peak dose regions are shown in Table 2. Striped GRID irradiation with 10 Gy nominal dose caused the greatest difference from the prediction model based on homogenous irradiation.

**Table 2.** Comparison between experimental and modelled A549 clonogenic survival for different peak doses. The estimates are presented as absolute percentage deviation and obtained by moving average along each cell flask in valley and peak regions.

| Peak dose (Gy) | Mean absolute percentage deviation GRID (valley; peak) |
|---|---|
| 1.6 | (8.3; 10.3) % |
| 4.1 | (4.0; 40.6) % |
| 8.2 | (3.5; 137.8) % |

## 4. Discussion

In the current study, by delivering a highly non-uniform dose distribution using a specially constructed grid collimator, the clonogenic survival of A549 cells following GRID irradiation was measured and compared to the survival predicted from open field irradiation data. EBT3 GFD and machine learning was used for pixel-by-pixel mapping of radiation dose and cell colonies, respectively.

The GRID dose distribution obtained from radiochromic EBT3 films was validated by numerical FLUKA MC simulations. The scored dose deposition maps from MC confirmed the finding from the film dosimetry of a lower peak dose relative to open field irradiation with the same nominal dose and a valley dose larger than 0 Gy in the shielded areas (see Figure 3). The 18% drop in dose from open field to peak GRID region is due to that the collimator will effectively attenuate diverging and scattered X-rays. The observed dose difference of 18% is however specific for the 5 mm striped fields. Bigger grid openings will yield less difference between peak dose and open dose deposition. Moreover, our detection of a low valley dose from scattered radiation in the collimated regions using GFD is in line with other *in vitro* studies (Mackonis *et al.*, 2007; Asur *et al.*, 2012; Peng *et al.*, 2017), where the valley dose is primarily dependent on the beam energy and modulation. For instance, Mackonis *et al.* (2007) measured approximately 6% valley dose relative to the prescribed dose for 6 MV striped intensity-modulated radiation therapy (IMRT) fields created by 0.75 cm openings and 2.25 cm MLC shieldings. Comparably with our collimated X-ray beam, Peng *et al.* (2017) generated a striped GRID pattern of periodicity of 5.0 mm spacing, which were characterized dosimetrically using radiochromic EBT3 film, where valley regions showed approximately 20% of prescribed dose. Other studies have reported an out-of-field dose <10% of the in-field dose (Trainor *et al.*, 2012; McGarry *et al.*, 2012).

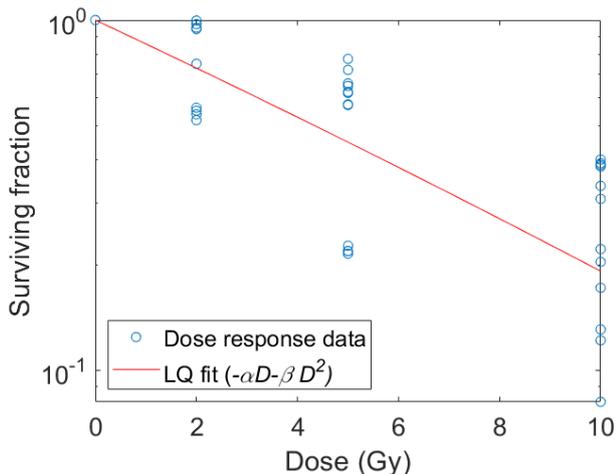

**Figure 4**. The surviving fraction, measured by deploying the image segmentation method (Arous *et al.*, 2022) on the colony assay, as a function of dose for A549 cells. The curve (red solid line) represents the linear-quadratic model-fit to processed colony data (blue open circles) for open field irradiation.

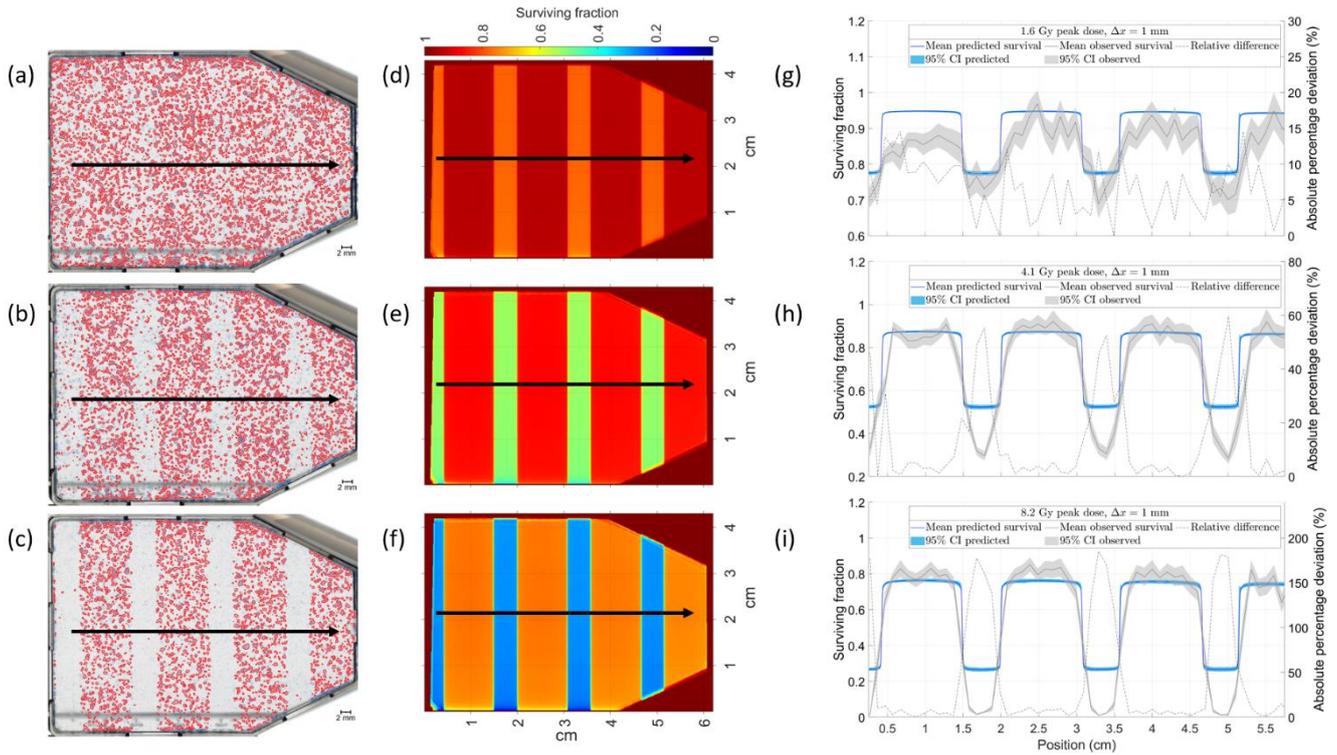

**Figure 5**. *Left panel*) Two-dimensional (2D) delineations suggested by the colony segmentation algorithm is presented in red for 2 (*a*), 5 (*b*) and 10 (*c*) Gy nominal dose of GRID irradiation. *Middle panel*) 2D surviving fraction distribution for 2 (*d*), 5 (*e*) and 10 (*f*) Gy nominal dose of GRID irradiation as predicted by the linear-quadratic (LQ) model. *Right panel*) Observed and predicted cell survival for 2 (*g*), 5 (*h*) and 10 (*i*) Gy nominal dose along the longitudinal direction of the cell flask. Observed survival profiles are presented with blue lines. The predicted survival profiles are presented with grey lines. Correspondingly coloured 95% CIs are presented with bands. The absolute percentage difference is presented as dotted line.

Several publications have stated that steep dose gradients associated with GRID IMRT (generally 6 MV X-ray beams) overall suggest significantly greater decrease in survival than expected in the collimated valley regions and unexpected increase in survival in the open peak regions for malignant melanoma (MM576), human prostate cancer (DU-145) and normal human fibroblast (AGO-1552) cells (Mackonis *et al.*, 2007; Butterworth *et al.*, 2011; Trainor *et al.*, 2012; McGarry *et al.*, 2012). Asur *et al.* (2012) also reported unexpectedly high cell kill in the valleys for confluent murine mammary carcinoma (SCK) and head and neck sarcoma (SCCVII) cells following exposure to a single dose of 10 Gy utilizing circular GRID opening of 12 mm diameter and a center-to-center distance of 18 mm (nearly 50:50 valley and peak exposure). In conclusion, these *in vitro* investigations of clonogenic survival demonstrated higher survival in the peaks and lower survival in the valleys.

We observed a significantly lower clonogenic survival of cells located in the peak regions compared to the predicted LQ-response, in contrast to previously published studies. The survival difference increased with increasing peak dose. We also observed a slightly lower survival level in adjacent valley cells compared to the predicted LQ-response. One can speculate whether this difference could be attributed to bystander cells killing which is absent from the classical LQ model formalism. Refinement of the classical LQ model with a bystander term have recently been reported (Peng *et al.*, 2017; Peng *et al.*, 2018). These predictive LQ models are based on the assumption that the generation of bystander signaling is linear in local dose. Therefore, this formulation increases the radiosensitivity contribution of the parameter α, which might explain the linearly increasing discrepancies in peak dose regions between the average observed and predicted cell survival levels (see Table 2). Ultimately, it can be seen that the standard LQ is not supported to be used to predict survival response for modulated X-ray fields, which will be further assessed in future GRID study.

## 5. Conclusion

A framework for detailed mapping of surviving A549 lung cancer cell colonies following GRID irradiation together with predicted survival levels based on data from open field irradiation has been presented. Our findings confirmed by MC simulations indicate that GRID irradiation causes lower doses in the peak regions, but higher doses in the valley regions than expected from the nominal dose. In addition, the observed cell survival was lower than the one predicted from the actual doses in both regions. The current developed methodology will aid future radiobiology investigations of GRID effects and cellular mechanisms causing the altered responses.

## Acknowledgements

This work was supported by the South-Eastern Norway Regional Health Authority (Project ID 2019050).

## Conflict of interest

The authors have no conflict of interest to report. The authors alone are responsible for the content and writing of this article.

## References

Arous D, Schrunner S, Hanson I, Frederike Jeppesen Edin N and Malinen E 2022 Principal component-based image segmentation: a new approach to outline in vitro cell colonies *Computer Methods in Biomechanics and Biomedical Engineering: Imaging & Visualization* 1-13

Asur R S, Sharma S, Chang C-W, Penagaricano J, Kommuru I M, Moros E G, Corry P M and Griffin R J 2012 Spatially fractionated radiation induces cytotoxicity and changes in gene expression in bystander and radiation adjacent murine carcinoma cells *Radiation research* **177** 751-65

Battistoni G, Boehlen T, Cerutti F, Chin P W, Esposito L S, Fassò A, Ferrari A, Lechner A, Empl A and Mairani A 2015 Overview of the FLUKA code *Annals of Nuclear Energy* **82** 10-8

Böhlen T, Cerutti F, Chin M, Fassò A, Ferrari A, Ortega P G, Mairani A, Sala P R, Smirnov G and Vlachoudis V 2014 The FLUKA code: developments and challenges for high energy and medical applications *Nuclear data sheets* **120** 211-4

Butterworth K T, McGarry C K, Trainor C, O'Sullivan J M, Hounsell A R and Prise K M 2011 Out-of-field cell survival following exposure to intensity-modulated radiation fields *International Journal of Radiation Oncology* Biology* Physics* **79** 1516-22

Devic S, Seuntjens J, Hegyi G, Podgorsak E B, Soares C G, Kirov A S, Ali I, Williamson J F and Elizondo A 2004 Dosimetric properties of improved GafChromic films for seven different digitizers *Medical physics* **31** 2392-401

Devic S, Tomic N and Lewis D 2016 Reference radiochromic film dosimetry: review of technical aspects *Physica Medica* **32** 541-56

Devic S, Wang Y Z, Tomic N and Podgorsak E B 2006 Sensitivity of linear CCD array based film scanners used for film dosimetry *Medical physics* **33** 3993-6

Ferrari A, Sala P R, Fasso A, Ranft J and Siegen U 2005 FLUKA: a multi-particle transport code. Citeseer)

Fowler J F 1989 The linear-quadratic formula and progress in fractionated radiotherapy *The British journal of radiology* **62** 679-94

Franken N A, Rodermond H M, Stap J, Haveman J and Van Bree C 2006 Clonogenic assay of cells in vitro *Nature protocols* **1** 2315-9

Gholizadeh Sendani N, Karimian A, Ferreira C and Alaei P 2018 Impact of region of interest size and location in Gafchromic film dosimetry *Medical physics* **45** 2329-36

Giard D J, Aaronson S A, Todaro G J, Arnstein P, Kersey J H, Dosik H and Parks W P 1973 In vitro cultivation of human tumors: establishment of cell lines derived from a series of solid tumors *J Natl Cancer Inst* **51** 1417-23

Ha J K, Zhang G, Naqvi S A, Regine W F and Yu C X 2006 Feasibility of delivering grid therapy using a multileaf collimator *Medical physics* **33** 76-82

Huhn J L, Regine W F, Valentino J P, Meigooni A S, Kudrimoti M and Mohiuddin M 2006 Spatially fractionated GRID radiation treatment of advanced neck disease associated with head and neck cancer *Technology in cancer research & treatment* **5** 607-12

Laissue J A, Blattmann H and Slatkin D N 2011 Alban Köhler (1874-1947): Inventor of grid therapy *Zeitschrift fur medizinische Physik* **22** 90-9

Mackonis E C, Suchowerska N, Zhang M, Ebert M, McKenzie D and Jackson M 2007 Cellular response to modulated radiation fields *Physics in Medicine & Biology* **52** 5469

McGarry C K, Butterworth K T, Trainor C, McMahon S J, O'Sullivan J M, Prise K M and Hounsell A R 2012 In-vitro investigation of out-of-field cell survival following the delivery of conformal, intensity-modulated radiation therapy (IMRT) and volumetric modulated arc therapy (VMAT) plans *Physics in Medicine & Biology* **57** 6635

Mohiuddin M, Curtis D L, Grizos W T and Komarnicky L 1990 Palliative treatment of advanced cancer using multiple nonconfluent pencil beam radiation: a pilot study *Cancer* **66** 114-8

Mohiuddin M, Fujita M, Regine W F, Megooni A S, Ibbott G S and Ahmed M M 1999 High-dose spatially-fractionated radiation (GRID): a new paradigm in the management of advanced cancers *International Journal of Radiation Oncology\* Biology\* Physics* **45** 721-7

Mohiuddin M, Stevens J H, Reiff J E, Huq M S and Suntharalingam N 1996 Spatially fractionated (GRID) radiation for palliative treatment of advanced cancer *Radiation Oncology Investigations: Clinical and Basic Research* **4** 41-7

Neuner G, Mohiuddin M M, Vander Walde N, Goloubeva O, Ha J, Cedric X Y and Regine W F 2012 High-dose spatially fractionated GRID radiation therapy (SFGRT): a comparison of treatment outcomes with Cerrobend vs. MLC SFGRT *International Journal of Radiation Oncology\* Biology\* Physics* **82** 1642-9

Niroomand-Rad A, Chiu-Tsao S-T, Grams M P, Lewis D F, Soares C G, Van Battum L J, Das I J, Trichter S, Kissick M W and Massillon-Jl G 2020 Report of AAPM task group 235 radiochromic film dosimetry: an update to TG-55 *Med. Phys* **47** 5986-6025

Peñagarícano J A, Moros E G, Ratanatharathorn V, Yan Y and Corry P 2010 Evaluation of spatially fractionated radiotherapy (GRID) and definitive chemoradiotherapy with curative intent for locally advanced squamous cell carcinoma of the head and neck: initial response rates and toxicity *International Journal of Radiation Oncology\* Biology\* Physics* **76** 1369-75

Peng V, Suchowerska N, Esteves A D S, Rogers L, Mackonis E C, Toohey J and McKenzie D R 2018 Models for the bystander effect in gradient radiation fields: Range and signalling type *Journal of theoretical biology* **455** 16-25

Peng V, Suchowerska N, Rogers L, Claridge Mackonis E, Oakes S and McKenzie D R 2017 Grid therapy using high definition multileaf collimators: realizing benefits of the bystander effect *Acta oncologica* **56** 1048-59

Trainor C, Butterworth K T, McGarry C K, McMahon S J, O'Sullivan J M, Hounsell A R and Prise K M 2012 DNA damage responses following exposure to modulated radiation fields

Vlachoudis V *Proc. Int. Conf. on Mathematics, Computational Methods & Reactor Physics (M&C 2009), Saratoga Springs, New York,2009),* vol. Series 176)